\documentclass{llncs}

\usepackage[T1]{fontenc}
\usepackage[utf8]{inputenc}
\usepackage[nocompress]{cite}
\usepackage[pdftex]{graphicx}
\usepackage{url}
\usepackage{array}
\usepackage[caption=false,font=footnotesize,labelfont=sf,textfont=sf]{subfig}
\usepackage{stfloats}
\usepackage[hidelinks,bookmarks,bookmarksdepth=3,bookmarksopen]{hyperref}

\usepackage[colorinlistoftodos, textwidth=2.0cm]{todonotes}
\newcounter{FRToDoCounter}

\begin{document}
\title{Towards a Viewpoint-specific Metamodel for Model-driven Development of\\Microservice Architecture}

\author{Florian Rademacher\inst{1}\orcidID{0000-0003-0784-9245} \and Jonas Sorgalla\inst{1}\orcidID{0000-0002-7532-7767} \and Sabine Sachweh\inst{1} \and Albert Z\"undorf\inst{2}}
\institute{Institute for the Digital Transformation of Application and Living Domains\\
	University of Applied Sciences and Arts Dortmund\\
	Otto-Hahn-Stra\ss{}e 23, 44227 Dortmund, Germany\\
	\email{\{florian.rademacher,jonas.sorgalla,sabine.sachweh\}@fh-dortmund.de}
	\and
	University of Kassel, Department of Computer Science and Electrical Engineering\\
	Software Engineering Research Group\\
	Wilhelmsh\"oher Allee 73, 34121 Kassel, Germany\\
	\email{zuendorf@uni-kassel.de}
}

\maketitle

\begin{abstract}
Microservice Architecture (MSA) is a service-based architectural style with a strong emphasis on high cohesion and loose coupling. It is commonly regarded as a descendant of Service-oriented Architecture (SOA) and thus might draw on existing findings of SOA research.

This paper presents a metamodel for Model-driven Development (MDD) of MSA, which is deduced from existing SOA modeling approaches, but also incorporates MSA-specific modeling concepts. It is divided into the three viewpoints Data, Service and Operation, each of which encapsulates concepts related to a certain aspect of MSA. The metamodel aims to support DevOps-based MSA development and automatic transformation of metamodel instances into MSA implementations.

\end{abstract}

\section{Introduction}
Microservice Architecture (MSA) \cite{Newman2015} is an architectural style for distributed software systems. It became apparent to practitioners and researchers in 2014 \cite{Francesco2017} and hence is relatively novel compared to its ancestor, Service-oriented Architecture (SOA) \cite{Erl2005,Pahl2016}. Both SOA and MSA leverage \textit{services} as architectural building blocks and consider them being software components that (i) are loosely coupled, i.e., minimize dependencies to other components; (ii) agree on \textit{contracts} as predefined specifications of communication relationships that enable service interaction; (iii) encapsulate reusable business or technical infrastructure logic; (iv) can be composed to coordinately accomplish coarse-grained tasks \cite{Newman2015,Erl2005}.

Besides this consistent understanding of service in SOA and MSA, microservices conceptually exhibit certain distinguishing characteristics to SOA services \cite{Rademacher2017,Francesco2017}. First, a microservice is responsible for providing exactly one distinct business or infrastructure functionality. Second, next to its logic's implementation, a microservice includes all technical artifacts necessary for deployment and execution, e.g., deployment descriptors and software frameworks. Third, MSA simplifies service interaction in that it (i) typically employs at most two communication protocols--one for synchronous and one for asynchronous interactions; (ii) prefers choreography over orchestration for architecture-internal service interaction; (iii) relies on lightweight API gateways
rather than full-fledged Enterprise Service Buses \cite{Erl2005} for interaction with external consumers. 

Based on the described characteristics and when compared to SOA, the adoption of MSA is commonly expected to increase (i) a system's adaptability; (ii) service quality and safety; (iii) productivity of development teams \cite{Rademacher2017}. On the other hand, SOA exhibits an extensive body of knowledge, resulting from more than a decade of research and practice \cite{Ameller2015}. Thus, the question arises, if and to what extent MSA might build upon existing findings of SOA research.


This paper contributes to answering this question in the area of Model-driven Development (MDD) \cite{RodriguesDaSilva2015}, whose application to SOA has been intensively studied \cite{Ameller2015}. Therefore, we present a \textit{metamodel} for MDD of MSA (MSA-MDD) that has been mainly deduced from existing approaches to SOA modeling. It is structured in different \textit{modeling viewpoints} \cite{OMG2014} to support DevOps-based MSA development \cite{Nadareishvili2016} by clustering modeling concepts for service developers and operators of a microservice team. The metamodel provides the basis for the subsequent implementation of \textit{modeling languages} \cite{RodriguesDaSilva2015} for MSA-MDD.

The remainder of the paper is organized as follows. Section~\ref{sec:metamodel} outlines the deduction of the metamodel from existing SOA modeling approaches and its viewpoints. Section~\ref{sec:implementation} describes the planned steps of the metamodel's implementation. Section~\ref{sec:related-work} presents related work and Section~\ref{sec:conclusion} concludes the paper.

\section{Metamodel Deduction and Viewpoints}\label{sec:metamodel}
This section elucidates the deduction process of the metamodel in Subsection~\ref{sub:concept-identification}. Subsections~\ref{sub:data-viewpoint}, \ref{sub:service-viewpoint} and \ref{sub:operation-viewpoint} cover the definitions of the metamodel's viewpoints.

\subsection{Deduction of Metamodel Concepts}\label{sub:concept-identification}
The metamodel's concepts were deduced from ten existing SOA modeling approaches. They were analyzed in full reading surveys with the goal to identify modeling concepts that are also applicable to MSA. While each analyzed approach comprises concepts spanning multiple phases of SOA engineering, from service identification to provision and operation, we were primarily interested in concepts for service design and operation modeling to support DevOps in MSA-MDD. Table~\ref{tab:soa-analysis} shows the analysis's steps and a summary of their results\footnote{The analysis and its results are described in detail in the preprint of a paper submitted to \hypersetup{pdfborder=1 1 1}\href{http://dsd-seaa2018.fit.cvut.cz/seaa/}{SEAA 2018}\hypersetup{pdfborder=0 0 0}. The preprint is available online \cite{RademacherSEAA2018Preprint}.}.


\begin{table}
	\centering
	\caption{Analysis steps and results for the deduction of the metamodel's concepts}
	\begin{tabular}{c|p{8.6cm}|>{\raggedright\arraybackslash}p{2.2cm}}
		\hline
		\bfseries Steps & \bfseries Description & \bfseries Results \\
		\hline\hline
		1, 2 & Extraction of modeling concepts from approach descriptions (step~1), and identification of their structure and approach-internal relationships to other concepts (step~2). & 434 extracted concepts \\
		\hline
		3 & Removal of concepts from the extracted set that (i) target phases of SOA engineering other than design and operation; (ii) enable modeling of sophisticated aspects of SOA like governance, policies or service behavior; (iii) were defined too imprecisely to deduce concepts' semantics. & 100 remaining concepts for service design and operation modeling \\
		\hline
		4, 5 & Bundling of semantically equivalent remaining concepts in concept clusters (step~4) and assessment of their applicability to MSA-MDD based on distinguishing characteristics between SOA and MSA \cite{Rademacher2017} (step~5). & 80 applicable concepts in 48 clusters \\
		\hline
	\end{tabular}
	\label{tab:soa-analysis}
\end{table}

Starting from the concept clusters yielded by analysis steps~4 and 5, the MSA-MDD metamodel was deduced and clustered in three distinct viewpoints. The viewpoint clustering was originally performed to facilitate modeling for typical stakeholders of DevOps-based MSA development \cite{Nadareishvili2016}, i.e., service developers and operators, by limiting the set of applicable modeling concepts based on stakeholders' tasks to eliminate conceptual clutter \cite{OMG2014}. Hence, the metamodel comprises a Service and an Operation viewpoint. However, our analysis yielded an additional Data viewpoint, whose concepts might be used by domain experts and service developers to express domain-specific data models of a microservice. 

We describe each viewpoint's structure below. To foster understanding, we state the analysis identifiers of defining SOA modeling approaches \cite{RademacherSEAA2018Preprint} when their terms are mentioned. Due to lack of space, we do not elucidate for every viewpoint concept on which applicable SOA modeling concepts it is based\footnote{This detailed information is provided as supplemental material under \url{https://fh.do/ecsa2018-sm}.}. 

\subsection{Data Viewpoint}\label{sub:data-viewpoint}
\begin{figure}
	\centering
	\includegraphics[scale=0.33,trim={1cm 1cm 1cm 1cm}]{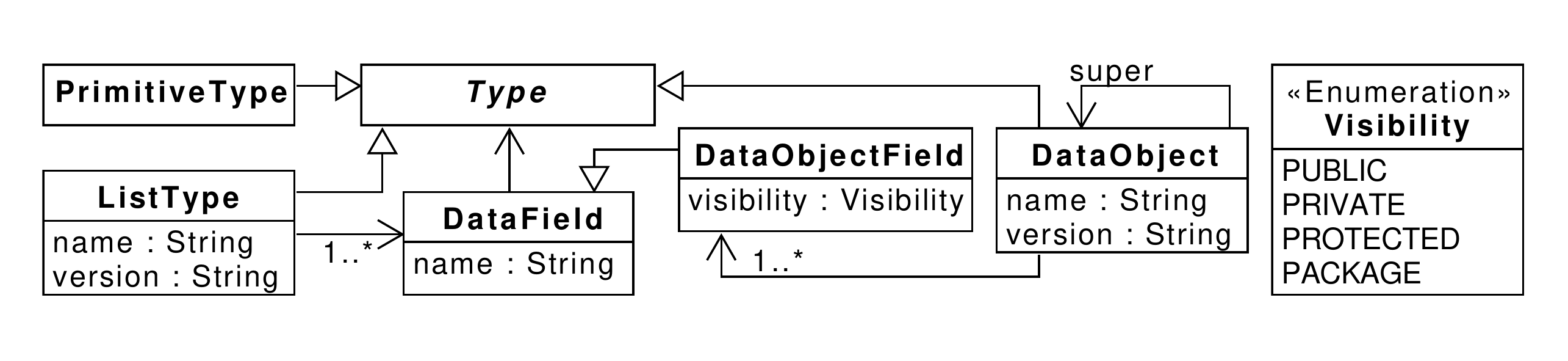}
	\caption{The Data viewpoint's modeling concepts and their relationships}
	\label{fig:data-viewpoint}
\end{figure}

The Data viewpoint, shown in Figure~\ref{fig:data-viewpoint}, comprises concepts to specify a microservice's information model (SOA modeling approaches A3, A7, A8 \cite{RademacherSEAA2018Preprint}).

Domain experts and service developers may express data structures for service interaction based on \texttt{ListTypes} and possibly nested, flat structured \texttt{Da\-ta\-Ob\-jects} \cite{Rademacher2015}, consisting of \texttt{DataFields} and \texttt{DataObjectFields}, respectively.

\subsection{Service Viewpoint}\label{sub:service-viewpoint}
\begin{figure}
	\makebox[\textwidth][c]{
		\includegraphics[scale=0.33,trim={1cm 1cm 1cm 1cm}]{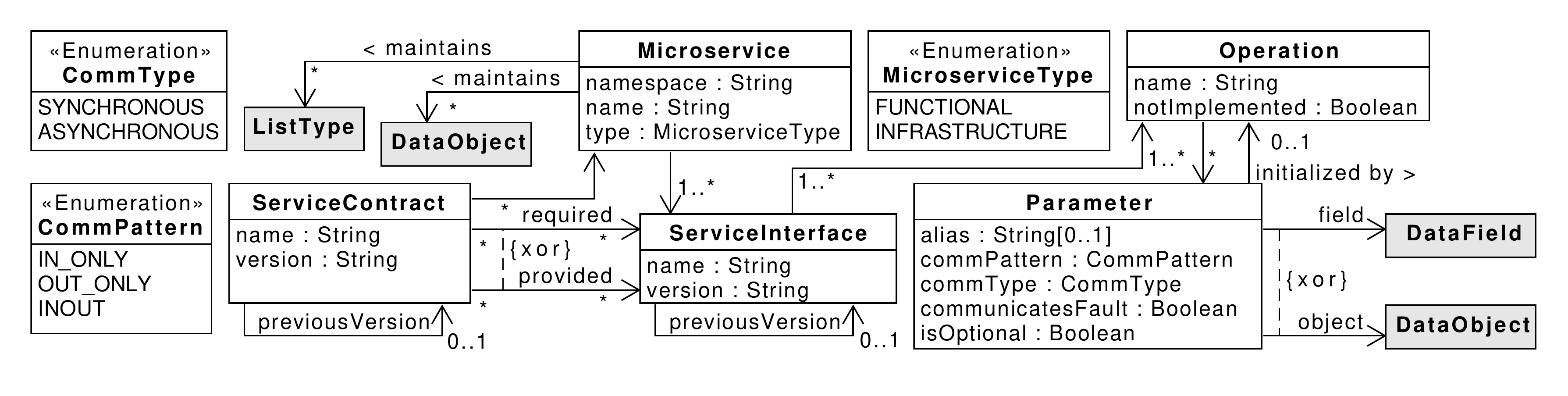}
	}
	\caption{The Service viewpoint's structure. Referenced Data concepts are colored gray.}
	\label{fig:service-viewpoint}
\end{figure}

The Service viewpoint, depicted in Figure~\ref{fig:service-viewpoint}, provides service developers with modeling concepts to specify microservices, interfaces and contracts. Therefore, it draws on data structures modeled with the Data viewpoint (cf. Subsection~\ref{sub:data-viewpoint}).

The Service viewpoint is centered around the \texttt{Microservice} concept. A microservice may either realize a functional, i.e., business-related, or an infrastructure capability \cite{Rademacher2017} specified by the \texttt{type} property. It also defines a \texttt{namespace} that encapsulates all associated concept instances.

A \texttt{Microservice} comprises at least one \texttt{ServiceInterface} with an \texttt{Op\-er\-a\-tion}. Operations expose service functions to consumers and enable service interaction (approaches A3 and A5 \cite{RademacherSEAA2018Preprint}). However, to support agile MSA development \cite{Nadareishvili2016} by specifying tentative, yet to be evolved, abstract function sets (A3 \cite{RademacherSEAA2018Preprint}), an operation may be marked as \texttt{not\-Im\-ple\-ment\-ed}.

An \texttt{Operation} may have \texttt{Parameters}. They specify the direction of information exchange based on WSDL-inspired \texttt{CommPatterns} (A5 \cite{RademacherSEAA2018Preprint}). Unidirectional input or output parameters exhibit the values \texttt{IN\_ONLY} or \texttt{OUT\_ONLY}. Bidirectional parameters are modeled as \texttt{INOUT}. The metamodel does not define message or event concepts (A4, A5, A8, A9 \cite{RademacherSEAA2018Preprint}) for asynchronous data exchange. Instead, the \texttt{CommType} enumeration specifies whether a parameter receives or provides data synchronously or asynchronously. 
The exchanged data's structure is determined based on the referenced Data concepts (cf. Subsection~\ref{sub:data-viewpoint}).

The metamodel does not define explicit concepts for sophisticated modeling of roles (A1, A2, A4, A5, A8, A9 \cite{RademacherSEAA2018Preprint}) and service interactions (A4, A5, A8--A10 \cite{RademacherSEAA2018Preprint}). The reason for omitting such concepts is MSA favoring choreography over orchestration \cite{Rademacher2017}. That is, microservice interactions boil down to service call chains between providers and consumers. Hence, the roles of provider and consumer in a service interaction are immediately clear due to their direct communication relationship. The Service viewpoint provides two ways to express service interactions resulting in choreography-driven call chains.

A Contract-driven Interaction Dependency (CDID) is expressed by a \texttt{Ser\-vice\-Con\-tract}. It bundles \texttt{ServiceInterfaces} provided and required by the contract's \texttt{Mi\-cro\-ser\-vice} in the sense of a ``binding agreement'' (A9 \cite{RademacherSEAA2018Preprint}).

A Parameter-driven Interaction Dependency (PDID) is given when a \texttt{Pa\-ram\-e\-ter} of one \texttt{Microservice} is \texttt{initialized} \texttt{by} an \texttt{Operation} of another microservice. In this case, an input parameter receives its value by invoking a functionality of another microservice.


\subsection{Operation Viewpoint}\label{sub:operation-viewpoint}
\begin{figure}
	\makebox[\textwidth][c]{
		\includegraphics[scale=0.33,trim={1cm 1cm 1cm 1cm}]{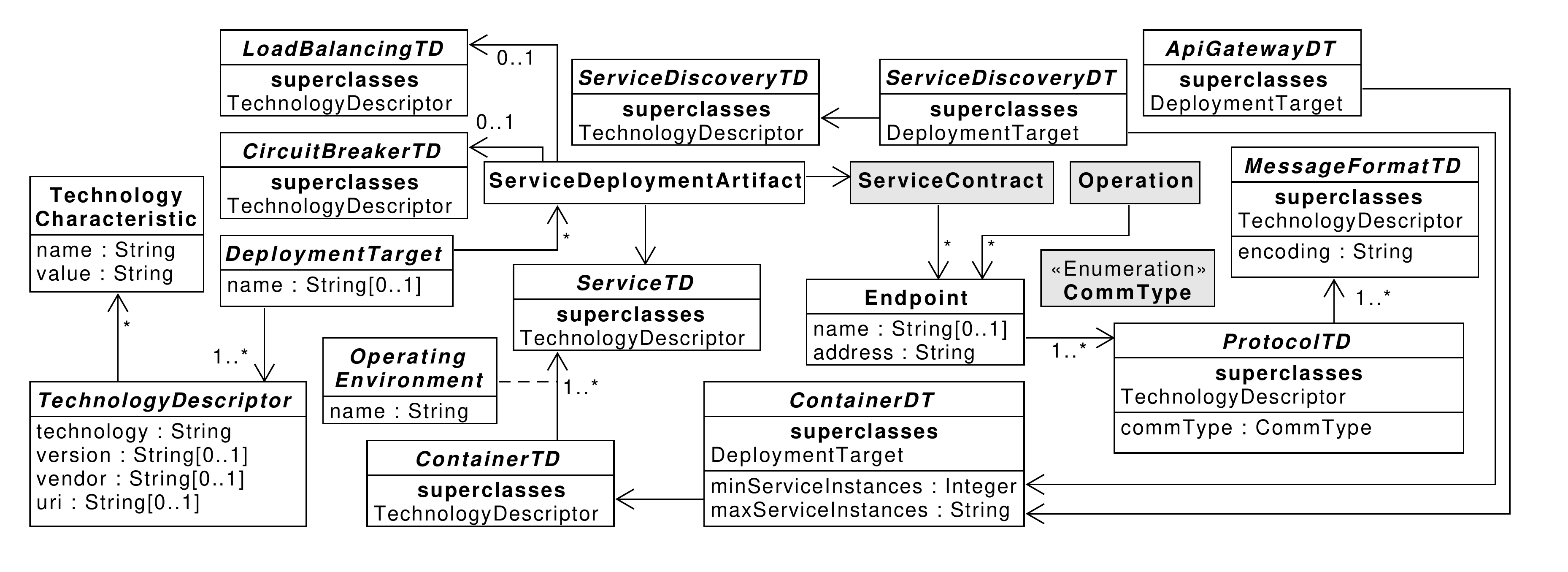}
	}
	\caption{The Operation viewpoint's structure with referenced Service concepts in gray. To increase clarity, concepts' superclasses are stated within compartments.}
	\label{fig:operation-viewpoint}
\end{figure}

The Operation viewpoint in Figure~\ref{fig:operation-viewpoint} clusters concepts used by service developers and operators to specify microservices' technology and deployment. Therefore, it refers to certain concepts of the Service viewpoint (cf. Subsection~\ref{sub:service-viewpoint}).

One of the central concepts of the viewpoint is \texttt{TechnologyDescriptor} (TD). It enables modeling of technologies for, e.g., service implementation, deployment or communication, and addresses MSA's inherent \textit{technology heterogeneity} \cite{Newman2015}.

\texttt{Endpoints} (A4 \cite{RademacherSEAA2018Preprint}) specify an \texttt{address}, communication protocols (\texttt{Pro\-to\-col\-TD}), e.g., synchronous REST or asynchronous AMQP\footnote{\url{https://www.amqp.org}}, and message formats (\texttt{Mes\-sage\-For\-mat\-TD}), e.g., XML or JSON, for a single \texttt{Operation} or a whole \texttt{Ser\-vice\-Con\-tract} of a \texttt{Microservice}.

Instead of a microservice, \texttt{ServiceContracts}, as means to specify service interactions (cf. Subsection~\ref{sub:service-viewpoint}), are bundled in \texttt{Ser\-vice\-De\-ploy\-ment\-Ar\-ti\-facts} (A10 \cite{RademacherSEAA2018Preprint}) and then deployed. This approach increases developer's flexibility because (i) interfaces of the microservice, i.e., those that are \texttt{provided} by the contract, may be selectively deployed; (ii) interactions of the deployed microservice may be deliberately limited to those with services \texttt{required} by the contract. Each artifact needs to have a \texttt{ServiceTD} assigned, which determines its implementation technology, e.g., Java in conjunction with Spring\footnote{\url{https://www.spring.io}}. Optionally, an artifact may exhibit service-internal infrastructure functionalities for load balancing (\texttt{LoadBalancingTD}) and fault mitigation (\texttt{CircuitBreakerTD}) \cite{Balalaie2015}.

A \texttt{ServiceDeploymentArtifact} is conceptually deployed by associating it with a \texttt{Con\-tain\-er\-DT} instance, i.e., the metamodel focuses on container-based MSA deployment \cite{Balalaie2015}. A container may specify the minimum and maximum service instances via its properties (A10 \cite{RademacherSEAA2018Preprint}). Moreover, each container has \texttt{ContainerTD} instances assigned, which, together with \texttt{ServiceTDs}, specify the \texttt{Op\-er\-at\-ing\-En\-vi\-ron\-ment} of a container. Its \texttt{name} points to a container image regarded as a combination of container technology and the service implementation technologies that need to be supported. For example, to deploy a Spring microservice in a Docker\footnote{\url{https://www.docker.com}} container, the \texttt{openjdk}\footnote{\url{https://hub.docker.com/_/openjdk}} Docker image may be employed. The assignment of more than one \texttt{ServiceTD} to a \texttt{ContainerTD} allows operators to proactively cope with MSA's technology heterogeneity by modeling containers that outright support more than one service implementation technology.

Finally, \texttt{ServiceDiscoveryDT} and \texttt{ApiGatewayDT} denote \texttt{De\-ploy\-ment\-Tar\-gets} to model service artifacts as internally or externally discoverable \cite{Balalaie2015}.

\section{Implementation of the Metamodel}\label{sec:implementation}
We expect the metamodel's design in its current form described in Section~\ref{sec:metamodel} to be mature enough to implement it for practical investigation and evaluation of MSA-MDD. Therefore, our implementation plan provides three steps, i.e., (S.1) viewpoint-specific language implementation; (S.2) integration of viewpoint languages; (S.3) deduction of a methodology for MSA-MDD.

Step~S.1 focuses on the implementation of a dedicated modeling language, including model validation and code generation, per metamodel viewpoint. This approach enables isolated development, testing and evaluation of each language. Thus, it facilitates iterative tailoring of a language to its respective stakeholders' concerns. For example, the Data viewpoint also addresses domain experts (cf. Subsection~\ref{sub:data-viewpoint}). A corresponding Data modeling language should hence be easy to learn and use for persons not necessarily having a technical background. In contrast, an Operation language for service developers and operators (cf. Subsection~\ref{sub:operation-viewpoint}) may incorporate technical terms and concise expressions known from programming or scripting languages for efficient textual modeling.

In step~S.2 we plan to integrate the Data and Service as well as the Service and Operation languages. The integration is necessary due to associations between modeling concepts from different viewpoints (cf. Figures~\ref{fig:service-viewpoint} and \ref{fig:operation-viewpoint}). Conceptually, an integrable language corresponds to a \textit{language component} \cite{Clark2015} and several tools exist to practically compose modeling languages. We are currently assessing the applicability of GEMOC Studio\footnote{\url{http://gemoc.org/gemoc-studio}}, MontiCore\footnote{\url{http://monticore.de}} and Melange\footnote{\url{http://melange.inria.fr}} for our use case. Therefore, we evaluate and compare the tools considering (i) degree of isolation of language development (higher is better); (ii) effort for code generator composition (lower is better); (iii) effort for adapting compositions to evolved languages (lower is better); (iv) range of supported types of concrete syntax \cite{RodriguesDaSilva2015}, i.e., textual, graphical or both (higher is better).

In step~S.3 we plan to deduce a methodology for MSA-MDD based on evaluating the languages and their integrations with different MSA teams.

\section{Related Work}\label{sec:related-work}
We present work related to languages and metamodels for MSA-MDD.

AjiL \cite{Sorgalla2017} is an MSA-MDD language based on the Eclipse Modeling Framework (EMF)\footnote{\url{https://www.eclipse.org/modeling/emf}}. It comprises a graphical editor for modeling microservices, their interfaces and exchanged data, as well as related technical infrastructure, e.g., service discoveries or load balancers (cf. Subsection~\ref{sub:service-viewpoint}). In addition, AjiL may generate executable Java code for each service. Compared to the proposed metamodel, AjiL exhibits several differences. First, it does not address different MSA-MDD stakeholders' concerns, e.g., by specialized languages or concrete syntaxes. Second, explicit technology modeling is not possible, e.g., service implementation is bound to Java, and protocols and message formats to REST and JSON. Third, AjiL does not support deployment modeling.

D\"ullmann and van Hoorn leverage an EMF-based metamodel to benchmark performance and resilience of microservices \cite{Dullmann2017}. Therefore, microservice stubs, including the necessary benchmark code, are generated from metamodel instances. The metamodel exhibits several similarities to the proposed one. First, interaction dependencies between microservices are modeled on the basis of operations (cf Subsection~\ref{sub:service-viewpoint}). Second, it comprises explicit concepts for service endpoints and operations (cf. Subsection~\ref{sub:operation-viewpoint}). Third, it covers basic deployment modeling. In contrast, however, the metamodel is not oriented towards various stakeholders, e.g., on the basis of viewpoints. It is further specialized in performance engineering and does not cover (i) explicit data modeling (cf. Subsection~\ref{sub:data-viewpoint}); (ii) specification of alternative technologies for service implementation and deployment besides Java and Docker; (iii) modeling of infrastructure components that are not mandatory for benchmarking, e.g., load balancers or circuit breakers.

\section{Conclusion and Future Work}\label{sec:conclusion}
In this paper we presented a metamodel for MSA-MDD. Its basic concepts were deduced from existing approaches to SOA modeling. Furthermore, it covers modeling of specific or frequent MSA concepts, e.g., API gateways or service-internal load balancers, and copes with microservices' inherent technology heterogeneity. To facilitate modeling for typical stakeholders in a microservice team, i.e., domain experts, service developers and operators, the metamodel was structured in three distinct viewpoints. They comprise only those concepts relevant to domain-specific Data, Service and MSA Operation modeling.

In future works we plan to implement the metamodel as a set of viewpoint-specific, integrated modeling languages (cf. Section~\ref{sec:implementation}). Next, we aim at evaluating the languages' degree of applicability in different MSA teams and refine the languages based on evaluation results. Eventually, we plan to deduce a methodology for MSA-MDD, which specifies structured usage of the modeling languages.


\bibliographystyle{spmpsci}
\bibliography{literature}
\end{document}